\documentstyle[preprint,aps,floats]{revtex}
\tightenlines
\def\Eslash{\not{\hbox{\kern-4pt E}}}

\begin{document}
\input{psfig.sty}

\preprint{
\font\fortssbx=cmssbx10 scaled \magstep2
\hbox to \hsize{
\includegraphics{uwlogo.ps}
\hskip.5in \raise.1in\hbox{\fortssbx University of Wisconsin - Madison}
\hfill$\vcenter{
\hbox{\bf MADPH-97-1000}
\hbox{\bf UCD-97-18}
\hbox{\bf hep-ph/9709366}
\hbox{September 1997}}$ }
}

\title{\vspace{.5in}
Sparticle Production in Electron-Photon Collisions}
\author{V. Barger$^*$, T. Han$^{*\dagger}$, and J. Kelly$^*$}
\address{$^*$Department of Physics, University of Wisconsin, Madison, WI
53706\\
$^\dagger$Department of Physics, University of California, Davis,
CA 95616}

\maketitle

\begin{abstract}
We explore the potential of electron-photon colliders to measure
fundamental
supersymmetry parameters via the processes $e\gamma \to \tilde e
\tilde\chi^0$
(selectron-neutralino) and $e\gamma\to \tilde\nu \tilde\chi^-$
(sneutrino-chargino). Given the $\tilde\chi^0$ and $\tilde\chi^-$ masses
from $e^+e^-$ and hadron collider studies, cross section ratios
$\sigma(\gamma_-)/\sigma(\gamma_+)$ for opposite photon helicities
determine
the $\tilde\nu_L$, $\tilde e_L$ and $\tilde e_R$ masses,
independent of the
sparticle branching fractions.
The difference $m_{\tilde\nu_L}^2-m_{\tilde e_L}^2$ measures
$M_W^2\cos2\beta$ in a model-independent way.
The $\tilde e_L$ and $\tilde e_R$ masses test the universality of
soft supersymmetry breaking scalar masses.
The cross section normalizations provide information about
the gaugino mixing parameters.
\end{abstract}
\thispagestyle{empty}
\newpage

A linear $e^+e^-$ collider with 0.5~TeV c.m.\ energy (expandable to
1.5~TeV)
and luminosity 50~fb$^{-1}$ per year (200~fb$^{-1}$ per year at
1.5~TeV) is of
special interest for the study of the properties of supersymmetric
particles \cite{susynlc,fengetal,msnu}.
In many unified models the lighter chargino ($\tilde\chi_1^\pm$)
and neutralinos ($\tilde\chi^0_{1,2}$)
are expected to be sufficiently light that they can be pair produced
at the Next Linear Collider (NLC). From the cross sections
for $e^+e^-\to\tilde\chi_1^+\tilde\chi_1^-$
and $\tilde\chi_1^0\tilde\chi_2^0$
and the kinematics of the
decays $\tilde\chi_1^\pm, \tilde\chi_2^0 \to \tilde\chi_1^0 f\bar f'$
to the lightest neutralino,
the masses of the $\tilde\chi_1^+, \, \tilde\chi_1^0$ and
$\tilde\chi_2^0$ would be determined,
along with valuable information
regarding their couplings.
Additionally, since one of two contributing Feynman graphs
involves slepton exchange, a determination or
upper bound on the slepton mass may be inferred\cite{fengetal,msnu}.
Experiments at the LHC may also measure the
$\tilde\chi_1^+, \, \tilde\chi_1^0$ masses and deduce
coupling information \cite{susylhc}.

In the minimal Supersymmetric Standard Model (MSSM),
the two gaugino mass parameters $M_1$ and $M_2$,
the Higgs mixing $\mu$, and the ratio of vacuum expectation
values, $\tan\beta=v_u/v_d$ fully
specify the gaugino masses and mixings. The sfermion masses
(such as $m_{\tilde e_L}, m_{\tilde e_R}\ \it etc.$)
involve additional soft SUSY breaking parameters.
In the minimal supergravity model
(mSUGRA)\cite{tata} with the assumption of universal soft parameters
at the GUT scale, the scalar mass $m_0$,
gaugino mass $m_{1/2}$, the trilinear coupling $A$ along with
$\tan\beta$ and the sign of $\mu$
are sufficient to determine all the physical
quantities at the electroweak scale. In the minimal
gauge-mediated SUSY breaking (GMSB) model\cite{dine}, the parameter set is
$\mu$, $\tan\beta$, the SUSY breaking vacuum expectation
value $F_X$ and the
messenger mass scale $M_X$; the effective SUSY breaking scale is
$\Lambda=F_X/M_X$.
Measuring the $\tilde\chi_1^+, \, \tilde\chi_1^0$ and
$\tilde\chi_2^0$ masses may be the first step toward deciphering
the nature of the SUSY model and especially for testing
the MSSM predictions for $M_1$ and $M_2$. However,
sfermion mass measurements will be essential to fully understand
the SUSY breaking mechanism.

In this Letter we consider the two processes
\begin{equation}
e\gamma\to\tilde\nu\tilde\chi^-
\qquad {\rm and} \qquad
e\gamma\to \tilde e\tilde\chi^0 \
\label{Process}
\end{equation}
for use in measuring the sneutrino ($\tilde\nu$)
and selectron ($\tilde e$) masses and their couplings.
We assume that the masses of the $\tilde\chi_1^{\pm}, \,
\tilde\chi_1^0$ and $\tilde\chi_2^0$
are known from experiments at the LHC or NLC.
At the LHC the mass reach for sleptons is only
about 250 GeV \cite{baerlhc}, due to the rather small electroweak
signal cross sections and large SM backgrounds.
In most SUSY
models, the sfermions
are heavier than the
lightest chargino and pair production of the scalar particles
$e^+e^-\to\tilde e^+\tilde e^-$ and $\tilde\nu\tilde\nu$
could be inaccessible at an $e^+e^-$ collider;
the kinematic thresholds for Eq.~(\ref{Process}) will then be lower
than those for sfermion pair production.
Even when scalar pair production is kinematically allowed
there is a $\beta^3$ suppression of the cross section near threshold,
and the event rates are correspondingly limited.
On the other hand, the
cross sections for  Eq.~(\ref{Process}) are proportional to
$\beta$ near threshold, so high production rates are achievable.
Thus the addition of
low energy laser beams to backscatter from the $e^\pm$ beams,
allowing high energy $e\gamma$ collisions \cite{egmcollider},
becomes very interesting for studying
sleptons.  The luminosity of the backscattered
photons is peaked not far below the $e^\pm$ beam energy,
$\langle E_\gamma\rangle \sim 0.83 \langle E_{e^\pm} \rangle$
for optimized laser energy,
so the c.m.\ energies (and luminosity)
available in $e\gamma$ collisions are comparable
to those of an $e^+e^-$ collider, $\langle
\sqrt{s_{e\gamma}}\rangle \sim 0.9
\langle \sqrt{s_{ee}} \rangle$ \cite{egmcollider}.
We find that high degrees of polarization
for $e^-$ and $\gamma$
beams are advantageous in the studies of the reactions
in~(\ref{Process}).

Due to approximate decoupling of Higgsinos
from the electron,
the cross sections for the processes
in~(\ref{Process})
are only large when the charginos and neutralinos
are mainly gaugino-like, namely
$\tilde\chi^\pm \sim \tilde W^\pm$ and
$\tilde\chi^0_1\sim \tilde B^0,\ \tilde\chi^0_2 \sim \tilde W^0$.
This situation corresponds to
$|\mu| \gg M_1, M_2$ in the MSSM. Fortunately gaugino-like
$\tilde\chi_1^\pm$, $\tilde\chi^0_1$ and $\tilde\chi^0_2$ are
highly favored theoretically
for two reasons: ({\it i}) the radiative electroweak symmetry
breaking in SUSY GUTs theories yields a large $|\mu|$ value if
$\tan\beta$ is bounded by the infrared fixed point solutions
for the top quark Yukawa coupling \cite{tanbeta,susygut};
({\it ii}) $\chi_1^0\sim\tilde B$ is strongly preferred for $\chi_1^0$ to be	
 a viable cold dark matter candidate \cite{tanbeta,susygut,dm}.
In the rest of our paper,
we will thus concentrate on this scenario, although we
comment later to what extent a
$\tilde\nu\tilde\chi^-$ signal with a small $\tilde W^-$ component
in $\tilde\chi^-$ can be detected.


\bigskip
\noindent\underline{Cross section formulae}

The process $e^-\gamma\to \tilde\nu  \tilde\chi^-$ proceeds via
$s$-channel
electron and $t$-channel chargino exchange; see Fig.~1(a). In the
$\tilde\chi^-$ coupling to $\tilde\nu_L$, the contribution from the
higgsino
is proportional to $m_e$ and thus can be neglected.
Consequently the scattering amplitude is proportional to the
wino fractions $V_{j1}$ of the matrix $V_{ji}$ that diagonalizes the
mass matrix (the first index $j$ labels the chargino mass eigenstate
$\tilde\chi^+_1$, $\tilde\chi^+_2$ and the second index $i$ refers to
the primordial gaugino and higgsino basis $\tilde W^\pm $, $\tilde
H^\pm$).
Further, the $\tilde\nu_L$ state fixes the incoming electron
chirality to be left-handed ``$-$", leaving just four independent
helicity amplitudes. The differential cross sections summed over
the chargino helicity  are
\begin{eqnarray}
&&\frac{d\sigma}{d\cos\theta}
(e^-_- \gamma_-  \to  \tilde\nu_L \tilde\chi_j^-) =
{\pi\alpha^2\over\sin^2\theta_W}\ {V_{j1}^2 \over s}\
\frac{r_{\tilde\chi}^2}{(1-\beta^2)}
\ \frac{\beta}{(1+\beta\cos\theta)^2 }\nonumber\\
&&\hspace{0.4in}\times\sum_{\lambda =\pm 1}
(1+\lambda\cos\theta) (1+\lambda\beta)
\left(\frac{\sqrt{1-\beta^2}}{r_{\tilde\chi}} -(1 + \lambda\beta)
\right)^2 ,
\label{charghel1}
\\
&&\hspace{0.0in}\frac{d\sigma}{d\cos\theta}(e^-_- \gamma_+  \to
\tilde\nu_L\tilde\chi_j^-) =
{\pi\alpha^2\over\sin^2\theta_W}\ {V_{j1}^2 \over s} \
\frac{ r_{\tilde\chi}^2 }{(1-\beta^2)}\ {2\beta^3
\sin^2\theta\ (1-\beta\cos\theta)\over (1+\beta\cos\theta)^2}.
\label{charghel2}
\end{eqnarray}
The subscripts on $e$ and $\gamma$ refer to the electron and photon
helicities. The angle $\theta$ specifies the chargino momentum
relative to the incoming electron direction in the c.m.\ frame,
$\beta=p/E$ is the chargino velocity in the c.~m.~frame,
and $r_{\tilde\chi}=m_{\tilde\chi}/\sqrt{s}$.
The $(\lambda_e,\lambda_\gamma)=(-,-)$ helicity amplitude is
$S$-wave near threshold so the cross section of Eq.~(\ref{charghel1}) is
proportional to $\beta$; the $(-,+)$ helicity
amplitude, which comes only from the $t$-channel diagram,
is $P$-wave near threshold so the cross section of
Eq.~(\ref{charghel2}) goes like $\beta^3$.
At high energies $\beta \to 1$, the helicity amplitudes develop
a well-known zero at $\cos\theta = 1$ \cite{robinn},
and the cross sections peak at  $\cos\theta = -1$.

Selectron-neutralino associated production $e\gamma\to\tilde
e\tilde\chi^0$ proceeds via $s$-channel electron
and $t$-channel selectron exchanges
\cite{selectron,kongoto,selectron2,ngwu};
see Fig.~1(b). Again, the contributions from higgsino components
($\tilde H_1^0, \, \tilde H_2^0$) of $\tilde\chi^0$
can be neglected and only the neutralino mixing
elements $Z_{j1}$ and $Z_{j2}$ enter. There are
eight independent helicity amplitudes to consider:
four for $\tilde e_L$ and four
for $\tilde e_R$. After summing over the neutralino helicities,
there are just four independent helicity cross sections as the
helicity of the
$\tilde e_R$ ($\tilde e_L$) matches that of the $e_R$ ($e_L$):
\begin{eqnarray}
&&\hspace{0.2in}\frac{d\sigma}{d\cos\theta}(e^-_+ \gamma_+\to\tilde
e_{R}^-
\tilde\chi^0_i) =\frac{d\sigma}{d\cos\theta}(e^-_- \gamma_-\to\tilde
e_{L}^-\tilde\chi^0_i)=
\pi\alpha^2 \ {2F_{i(L,R)}^2\ \over s} \
\frac{r^2_{\tilde e}}{(1-\beta^2)}
\nonumber\\
&&\hspace{0.4in}
\times \frac{\beta}{(1+\beta\cos\theta)^2} \
\sum_{\lambda =\pm 1} (1+\lambda\cos\theta) (1+\lambda\beta)^2
\left(\frac{\sqrt{1-\beta^2} }{r_{\tilde e}} - (1+\lambda\beta)\right),
\label{neuthel1}\\
&&\hspace{0.2in}\frac{d\sigma}{d\cos\theta}(e^-_+ \gamma_-\to\tilde
e_{R}^-
\tilde\chi^0_i) = \frac{d\sigma}{d\cos\theta}(e^-_- \gamma_+\to\tilde
e_{L}^-\tilde\chi^0_i)=\nonumber\\
&&\hspace{0.65in}
\pi\alpha^2\ {2F_{i(L,R)}^2 \over s}\ \frac{r^2_{\tilde e}}{(1-\beta^2)}
\ \frac{2\beta^3\sin^2\theta}{ (1+\beta\cos\theta)^2 }
\left(\frac{\sqrt{1-\beta^2}}{r_{\tilde e}} -
(1-\beta\cos\theta) \right).\label{neuthel2}
\end{eqnarray}
The $F_{iL}$ $(F_{iR})$ for $\tilde e_L$ $(\tilde e_R)$
are effective couplings given by
\begin{eqnarray}
F_{iL} = \frac{1}{2}\left[\frac{Z_{i1}}{\cos\theta_W}
+\frac{Z_{i2}}{\sin\theta_W}\right], \quad
F_{iR} = \frac{-Z_{i1}^{*}}{\cos\theta_W}.
\label{lrcouplings}
\end{eqnarray}
Here the $Z_{ji}$ are the elements of matrices that diagonalize
the neutralino mass matrix (the first index $j$ labels the neutralino
mass eigenstate $\tilde\chi_j^0$, $j=1...4$, and the second index
$i=1,2$ refers to the primordial gaugino and higgsino basis
($\tilde B^0,\, \tilde W^0,\, \tilde H_1^0,\,\tilde H_2^0$)).
The angle $\theta$ specifies the direction of the selectron
with respect to the direction of the incoming electron in the c.m.\
frame,
$\beta$ is the velocity of the selectron, and
$r_{\tilde e}=m_{\tilde e} /\sqrt{s}$. Again,
at high energies $\beta \to 1$, the cross sections develop
a zero at $\cos\theta = 1$,
and peak at  $\cos\theta = -1$.

\bigskip
\noindent\underline{Parameters}

For our illustrations we choose the
chargino/neutralino masses (in GeV)
\begin{equation}
m_{\tilde\chi_1^0} = 65 \qquad m_{\tilde\chi_2^0} = 136 \qquad
m_{\tilde\chi_1^\pm} = 136 \qquad m_{\tilde\chi_2^\pm} = 431,
\end{equation}
and the mixing matrix elements
\begin{equation}
V_{11} = 0.98 \qquad V_{21} = -0.18
\end{equation}
for the charginos and
\begin{equation}
Z_{11} = 0.98 \qquad Z_{12} = -0.15 \qquad Z_{21} = -0.18 \qquad
Z_{22} = -0.96
\end{equation}
for the neutralinos.  These parameters correspond to the following MSSM
parameters at the weak scale
\begin{equation}
M_1 = 62 {\rm\ GeV},\quad M_2 = 127\ {\rm GeV}, \quad
\mu = 427\rm\ GeV\ , \quad \tan\beta=1.8\ ,
\label{masstanb}
\end{equation}
where $\tan\beta$ is at the infrared
fixed point value\cite{tanbeta} and the convention
for sign$(\mu)$ follows Ref.~\cite{tanbeta}.
For slepton masses, we choose (in GeV)
\begin{eqnarray}
m_{\tilde e_L}=320 \qquad m_{\tilde e_R}=307\qquad  m_{\tilde\nu_L}=315 ,
\end{eqnarray}
where $m_{\tilde e_L}$ and $m_{\tilde e_R}$ are independent
parameters in MSSM. Our choices for the gaugino and slepton masses
are consistent with
renormalization group evolution \cite{susygut}
to the electroweak scale,
with the following universal mSUGRA parameters
\begin{eqnarray}
 m_0 = 300\rm\ GeV\,,\qquad
 m_{1/2} = 150\rm\ GeV\,,\qquad
 A = 0 \,.
\label{gutmass}
\end{eqnarray}

\bigskip
\noindent\underline{$e^-\gamma\to \tilde\nu_L\tilde\chi^-$ cross section}

The total cross sections are shown
in Fig.~2(a) for $\tilde\nu_L\tilde\chi_1^-$
production and in Fig.~2(b) for $\tilde\nu_L\tilde\chi_2^-$
versus $\sqrt s$.
The peak cross section is about 1.5 picobarns for
$\tilde\nu_L\tilde\chi_1^-$, and is smaller
for $\tilde\nu_L\tilde\chi_2^-$
partially due to the smaller $V_{21}$ coupling and partially due to the
energy-dependent factor.
We have numerically checked that our calculations agree with
Ref.~\cite{robinn} for his particular choice of
$m_{\tilde\chi_1^-}=m_{\tilde\nu_L}$ and couplings.

For realistic predictions we convolute these subprocess cross sections
with the appropriate backscattered photon spectrum
\cite{egmcollider}; these results are shown by
the lower curves in Fig.~\ref{charginoxsecfig}.
The effect of the convolution is to decrease the cross sections by
about a factor of two.
In our illustrations we choose a pre-scattering laser beam energy of
$\omega_0=1.26$ eV.  For $E_e>250$ GeV a lower $\omega_0$ would
be needed to avoid electron pair
production at the backscattering stage.  We assume a polarization
$P_{e(L,R)}=0.9$
of the non-scattered electron beam, a mean helicity $\lambda=\pm
0.4$ of the pre-scattered electron beam, and a fully polarized
$P_c=\pm 1$ pre-scattered photon beam.  We select $\lambda
P_c$ to be negative to have a
relatively monochromatic spectrum of $\sqrt{s_{e \gamma}}$, peaked
close to $0.9 \sqrt{s_{ee}}$.
The individual signs of $\lambda$ and $P_c$ are chosen to
illustrate the helicity cross sections.

\noindent\underline{$e^-\gamma\to \tilde e^-\tilde\chi^0$ cross section}

The total cross
sections for $\tilde e^-\tilde\chi^0$ production
versus  $e\gamma$ c.m.\ energy are illustrated
in Fig.~\ref{neutralinoxsecfig2},
where the four panels present results for
$\tilde e_L$ and $\tilde e_R$
and for $\tilde\chi_1^0$ and  $\tilde\chi_2^0$, in all combinations.
Due to the stronger diagonal couplings $Z_{ii}$,
the cross sections for
$\tilde e_L \tilde\chi_2^0$
(mainly $\tilde e_L \tilde W^0$) and
$\tilde e_R \tilde\chi_1^0$
(mainly $\tilde e_R \tilde B^0$)
are significantly larger (see Eq.~\ref{lrcouplings});
the weaker neutral current couplings and the more massive
scalar propagator in selectron production make the cross
sections smaller than for sneutrino production.  The lower
curves in Fig.~\ref{neutralinoxsecfig2} show the effects of convolution
with the backscattered laser photon spectrum for the machine
parameters detailed previously.  Again the effect is to decrease the
cross sections by about
a factor of two. We numerically compared our
convolution results
with Fig.~1 of Ref.~\cite{kongoto} and found
excellent agreement.

\bigskip
\noindent\underline{Signal Final States and Backgrounds}

The decays of the sparticles in these reactions generally give clean
signals: large missing energy ($\Eslash$), energetic charged
leptons or jets from light quarks.
Table~\ref{msugra-bf} gives predicted branching
fractions \cite{isasugra}
for the mSUGRA example discussed earlier.
Based on the predicted cross
sections in Figs.~2 and 3, we concentrate on the three leading
channels
\begin{equation}
e^- \gamma \to \tilde\nu_L \tilde\chi_1^-, \quad
e^- \gamma \to \tilde e_R \tilde\chi_1^0\quad {\rm and} \quad
e^- \gamma \to \tilde e_L \tilde\chi_2^0 \ .
\label{channels}
\end{equation}
The cross section for
$\tilde\nu_L \tilde\chi_1^-$ production is of
$\cal O$(100-1000 fb), while the cross sections for
$\tilde e_R \tilde\chi_1^0$ and $\tilde e_L \tilde\chi_2^0$
are typically of $\cal O$(10-100 fb).

\begin{table}[t]
\caption{\label{msugra-bf} Sparticle decay modes and branching
fractions for the representative parameter choice. Here $q$
generically denotes a quark and  $\ell=e$ or $\mu$;
fermion-antifermion pairs $f\bar f'$ with net charge $-1$ ($0$) are
denoted by $C^-$ ($N^0$).}
\tabcolsep=1in
\begin{tabular}{cc}
Decay Modes& Branching Fraction (\%)\\
\hline
$\tilde\chi_1^-\to \tilde\chi_1^0 C^-(f\bar f')$& 62
($\tilde\chi_1^0 q\bar q'$), 25
($\tilde\chi_1^0
\ell^-\nu$)\\
\hline
$\tilde\chi_2^0\to \tilde\chi_1^0 N^0(f\bar f)$& 66
($\tilde\chi_1^0 q\bar q$), 20
($\tilde\chi_1^0\nu\bar\nu$), 9 ($\tilde\chi_1^0\ell^+\ell^-$) \\
\hline
$\tilde e_R\to \tilde\chi_1^0 e^-$& 97 \\
\hline
$\tilde e_L\to \tilde\chi_1^-\nu$, $\tilde e_L\to \tilde\chi_2^0
e^-$, $\tilde
e_L\to \tilde\chi_1^0 e^-$ & 59, 36, 5 \\
\hline
$\tilde\nu_L\to\tilde\chi_1^+ e^-$, $\tilde\nu_L \to \tilde\chi_2^0\nu$,
$\tilde\nu_L\to \tilde\chi_1^0\nu$& 59, 23, 18
\end{tabular}
\end{table}

\begin{table}
\bigskip
\caption{\label{process-bf} Sparticle production and decay
in $e^-\gamma$ collisions. Branching fractions based on
Table I are given in the parentheses.
Here $\Eslash$ denotes missing energy resulting from
$\tilde\chi_1^0$ and $\nu$ final state; $C^-$ ($N^0$)
denotes a fermion-antifermion pair of charge $-1$ ($0$).}
\begin{center}
\begin{tabular}{cc}
Process & Final State \& Branching Fraction\\
\hline
$e^- \gamma \to \tilde\nu_L \tilde\chi_1^- \to$
& $C^- \ \Eslash$ \ (18\%),\quad $ C^- \ N^0\ \Eslash $\  (23\%),\quad
  $ C^- \ C^+  e^- \ \Eslash$\ (59\%) \\
\hline
$e^- \gamma \to \tilde e_R \tilde\chi_1^0 \to$
& \ $ e^- \Eslash$ \ (100\%) \\
\hline
$e^- \gamma \to \tilde e_L \tilde\chi_2^0 \to$
& $ N^0 e^-\ \Eslash$\ (5\%),\quad $ N^0 \ N^0 e^-\ \Eslash$\  (36\%),
\quad $C^- \ N^0 \ \Eslash$\ (59\%)
\end{tabular}
\end{center}
\end{table}

In Table~\ref{process-bf}, we list observable final states
for these three processes. We have calculated the SM backgrounds,
which are presented in Table~\ref{bckgrnd}.
$W^-$ ($Z$) in the backgrounds will give the same final state as
$C^-$ ($N^0$)
in the signal table \ref{bckgrnd}, although the latter will often
be non-resonant $f\bar f'$ pairs. Generally speaking,
the multiple $f\bar f'$ signals for $\tilde\nu_L \tilde\chi_1^-$
and $\tilde e_L \tilde\chi_2^0$ have favorable
signal/background ratios, especially at
$\sqrt s_{e\gamma} \sim$ 0.5 TeV.
Given the sizeable signal cross
sections and the distinctive kinematical characteristics
from the heavy sparticle decays, such final states
do not have severe SM backgrounds.
However, some care needs to be taken for
the $\tilde e_R^{} \tilde\chi_1^0$ signal,
which decays exclusively to $e^-$ plus $\Eslash$.
The cross section for the SM background
$e^- \gamma \to e^- \nu \bar \nu$ (mainly from $e^- \gamma \to W^-
\nu$) is large, on the
order of a few picobarns. Detailed
analysis \cite{selectron2} shows that by
making use of a polarized $e^-_R$ beam and
kinematical cuts,
the $\tilde e^{}_R$ signal can be separated
from the SM background.  In our subsequent discussion
we will simply assume a 30\% efficiency associated with signal
branching ratios and
acceptance cuts for each channel
in (\ref{channels}), and assume no significant backgrounds remain after
the appropriate selection cuts have been implemented.

\begin{table}[t]

\caption{\label{bckgrnd} Total cross sections (in fb) for
missing energy plus vector bosons as Standard Model backgrounds
to SUSY signals in $e^-\gamma$ collisions at c.m.\ energies
$\protect \sqrt s_{e\gamma} $ = 0.5, 1.0 and 1.5 TeV.}
\begin{tabular}{cccc}
$e\gamma \to X$
& $\sqrt s_{e\gamma}=0.5$ & 1.0  & 1.5 \\ \hline
$X=W^- \nu_e$ & $\sigma=4.2\times 10^4$ fb & 4.8$\times 10^4$ &
4.9$\times 10^4$ \\ \hline
$e^-\nu \bar \nu$ & 6.5$\times 10^3$ &6.3$\times 10^3$ &
6.3$\times 10^3$ \\ \hline
$ W^- Z\nu_e $ & 210 & 720 & 1.1 $\times 10^3$ \\ \hline
$ Z e^- \nu\bar \nu $ & 23 & 79 & 120 \\ \hline
$ W^-W^+ e^- \nu\bar \nu $ & 0.62 & 8.6 & 21 \\  \hline
$ ZZ e^- \nu\bar \nu$ & 3 $\times 10^{-2}$ & 0.7 & 2 \\
\end{tabular}
\end{table}

Before we proceed, a few comments are in order.
First, given the negligibly small background
to the signal $\tilde\nu_L \tilde\chi_1^-$ in $C^- C^+ e^- \Eslash$
at $\sqrt s_{e\gamma} \sim$ 0.5 TeV,
it should be possible to measure even a small signal
for this channel, allowing a probe of a
small gaugino component in $V_{11}$.
The maximum value of the cross section for $\tilde\nu_L \tilde\chi_1^-$
is about $600 V_{11}^2$ fb, with a realistic
photon spectrum. If a cross section of 20 fb is needed
for a clear signal, a sensitivity down to
$V_{11}\sim 0.2$ would be feasible.
Second, although we have not discussed the channel
$e^- \gamma \to \tilde e_L \tilde\chi_1^0$ because
it has a smaller cross section for our parameter choices,
there is a cross section complementarity
between $\tilde e_L \tilde\chi_1^0$
and $\tilde e_L \tilde\chi_2^0$, depending on which
$\tilde\chi_j^0$ has a larger $\tilde W^0$ component.
Finally, the cross section normalization for
$\tilde e_R^{} \tilde\chi_1^0$ via the exclusive decay
$\tilde e_R \to \tilde\chi_1^0 e^-$ will directly measure
the neutralino mixing parameter $Z_{11}$ (or $F_{1R}^{}$).
On the other hand, one would have to measure the total
cross sections through all the decay channels for
$\tilde \nu_L $ and $\tilde e_L$ to determine the other mixing
parameters $Z_{ji}$ and $V_{j1}$.

\begin{figure}
\centerline{\psfig{file=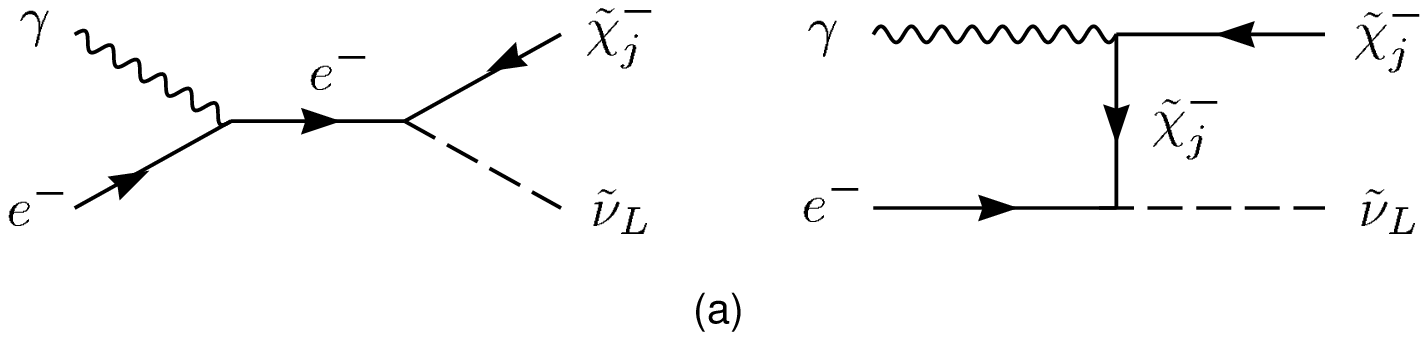,angle=0,width=10.5cm}}
\smallskip
\centerline{\psfig{file=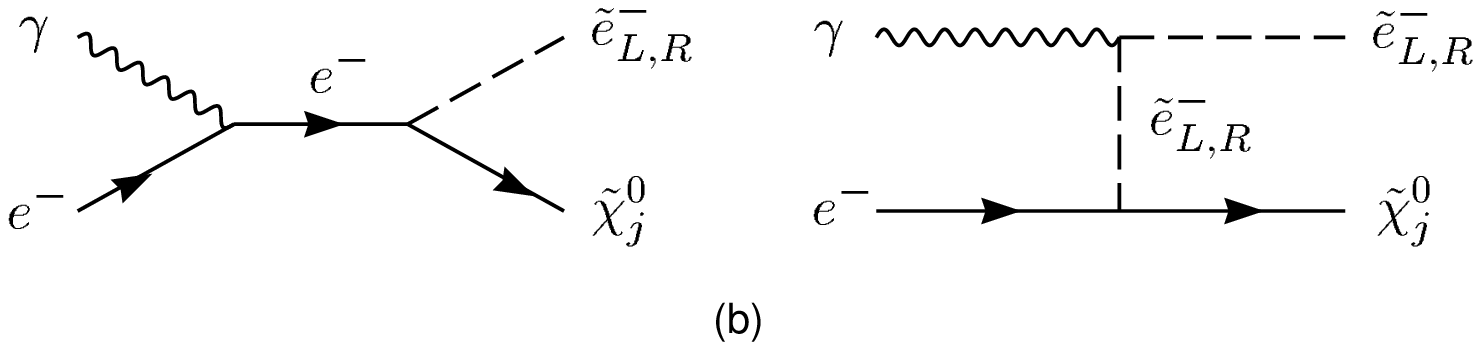,angle=0,width=10.5cm}}

\caption[]{Feynman graphs for the processes
(a)~$e^-\gamma\to\tilde\nu\tilde\chi^-$ and
(b)~$e^-\gamma\to \tilde e^-\tilde\chi^0$.
\label{graphs}}
\end{figure}

\begin{figure}
\centerline{\psfig{file=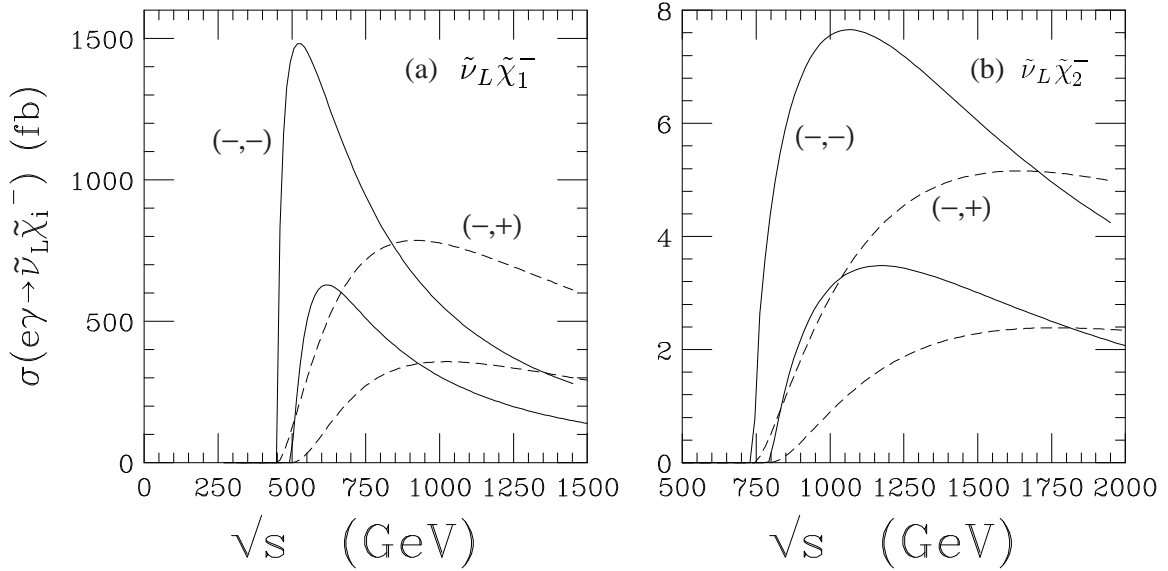,angle=90,width=15.5cm}}
\bigskip
\caption[]{The upper two curves show the total cross section (in fb)
for $e^-\gamma\to\tilde\nu
\tilde\chi^-$ versus $\protect\sqrt{s_{e\gamma}}$ (in GeV)
for the SUSY and machine parameters given in the text:
(a)~$\tilde\nu\tilde\chi_1^-$; (b)~$\tilde\nu
\tilde\chi_2^-$.  The solid curves represent $e$, $\gamma$
helicities $(-,-)$ and the dashed curves $(-,+)$.  The lower two curves
are corresponding results convoluted with the backscattered
photon spectrum versus $\protect\sqrt{s_{ee}}$.}
\label{charginoxsecfig}
\end{figure}

\begin{figure}
\centerline{\psfig{file=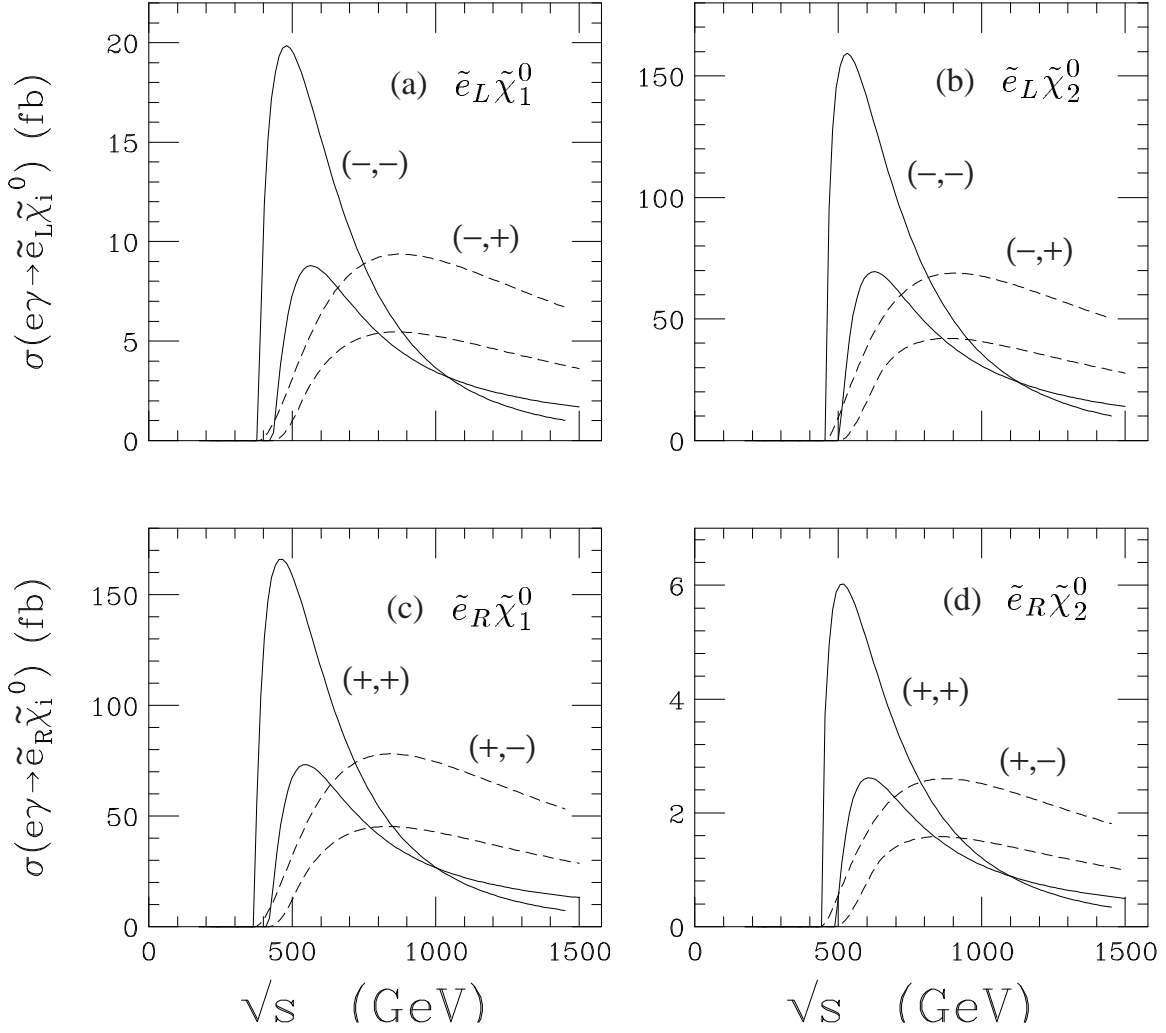,angle=90,width=15.5cm}}
\bigskip
\caption[]{The upper two curves show the total cross section (in fb)
for $e^-\gamma\to\tilde e\tilde\chi^0$
versus $\sqrt{s_{e\gamma}}$ (in GeV)
for the SUSY and machine parameters given in the text:
(a)~$\tilde e_L\tilde\chi_1^0$; (b)~$
\tilde e_L\tilde\chi_2^0$; (c)~$\tilde e_R\tilde\chi_1^0$; (d)~$\tilde
e_R\tilde\chi_2^0$.  The solid curves represent $e$, $\gamma$
helicities $(-,-)$ for (a), (b) and $(+,+)$ for (c), (d).  The dashed
curves represent helicities $(-,+)$ for (a), (b) and $(+,-)$ for (c),
(d).  The lower two curves
are corresponding results, convoluted with the backscattered
photon spectrum, versus $\sqrt{s_{ee}}$.
\label{neutralinoxsecfig2}}
\end{figure}

The preceeding signal and background discussions
are based on the mSUGRA scenario.
For gauge-mediated SUSY breaking  models
the signals from (\ref{graphs}) may be more spectacular.
For instance, if $\tilde\chi_1^0$ is the next-to-LSP (NLSP),
it will decay to a LSP gravitino ($\tilde G$)
plus a photon, resulting in an isolated photon
plus $\Eslash$ signal if the $\tilde\chi_1^0$ decay length is short
\cite{ngwu}. In other GMSB models in which
the NLSP is a right-handed slepton, the signal from
$\tilde\chi_1^0 \to l^\pm \tilde l^\mp_R$ is also spectacular.  Since
signals from GMSB models should be easily detectable at
the $e^+e^-$ NLC \cite{gmsusy} and the Tevatron\cite{baeretal},
we do not discuss such possibilites further in $e\gamma$ processes.

\bigskip
\noindent\underline{Slepton mass determination:
cross section ratio and energy endpoint measurement}

\begin{figure}
\centerline{\psfig{file=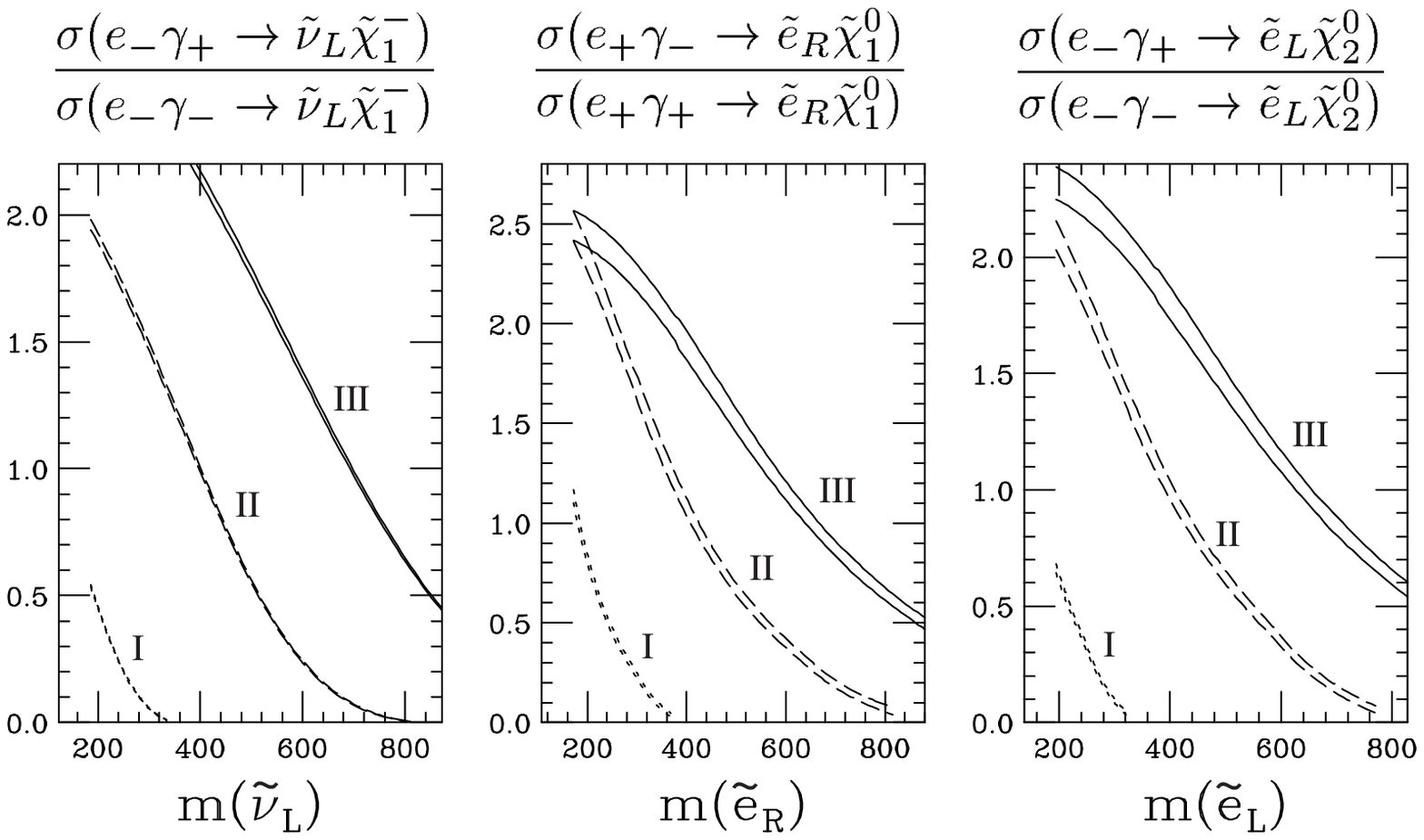,angle=90,width=15.5cm}}
\bigskip
\caption[]{The ratio of total cross sections versus
slepton mass (in GeV):\\
(a)~$\sigma(e_- \gamma_+  \to
\tilde\nu_L\tilde\chi_1^-)/\sigma(e_- \gamma_-
\to\tilde\nu_L\tilde\chi_1^-)$
 vs $m_{\tilde\nu_L}$; (b)~$\sigma(e_+ \gamma_-  \to
\tilde e_R\tilde\chi_1^0)/\sigma(e_+ \gamma_+
\to\tilde e_R\tilde\chi_1^0)$ vs $m_{\tilde e_R}$; (c)~$\sigma(e_-
\gamma_+  \to
\tilde e_L\tilde\chi_2^0)/\sigma(e_- \gamma_-
\to\tilde e_L\tilde\chi_2^0)$ vs $m_{\tilde e_L}$.
Results for three colliders are presented:
(I)~$\sqrt{s_{ee}}=0.5$ TeV, ${\cal L}_{e\gamma}= 25$ fb$^{-1}$;
(II)  $1$
TeV, $50$ fb$^{-1}$; (III)  $1.5$ TeV, $100$ fb$^{-1}$.  For each
collider
the upper and lower curves show the $\pm 1\sigma$ values for the ratio.
The backscattered photon spectrum and the electron polarization are
included in the calculations (see text).
\label{ratiofig}}
\end{figure}

In any of the $e\gamma$ processes under consideration the ratio of
the cross sections for the two photon helicities provides
a direct measure of the slepton mass if the associated
neutralino or chargino mass is already known from
$e^+e^-$ or $pp$ collider experiments. These ratios are independent
of the
cross section normalization factors and the final state branching
fractions.
Figure~\ref{ratiofig} shows the ratios
for these three leading channels:
(a) $\tilde\nu_L \tilde\chi_1^-$;
(b) $\tilde e_R \tilde\chi_1^0$;
(c) $\tilde e_L \tilde\chi_2^0 $.
The three pairs of bands on each panel correspond to the different
energy choices of $\sqrt{s_{ee}} = 0.5$, $1$, and $1.5$~TeV at
integrated luminosities of 25, 50, and 100~${\rm fb}^{-1}$, respectively,
convoluted with the backscattered photon spectrum.
Each pair of bands represents the $\pm 1\sigma$ error bounds on the
ratio.
We have included a $30\%$ efficiency factor for signal
identification with rejection of the SM backgrounds. From the figures,
we find that the $\pm 1\sigma$ uncertainties
determined by the cross section ratios translate into mass
uncertainties  of
roughly 2~GeV in channel (a),
and $6$, $30$ and $35$~GeV in channels (b)
and (c) for the respective energies given above.
%

Another way to measure the slepton mass is through the endpoint
of energy distribution in the two-body decay $\tilde e_R\to
\tilde\chi_1^0
e^- $\cite{susynlc}.
If the maximum (minimum) energy of the electron in
the lab frame is $E_+$ ($E_-$), then the selectron mass
can be determined from
\begin{equation}
m_{\tilde e_R} \ =\
\frac{s \sqrt{E_+ E_-} }{\sqrt s (E_+ + E_-) - 2E_+ E_- },
\end{equation}
and the LSP mass is given by
\begin{eqnarray}
m_{\tilde\chi_1^0}^2\ =\ m_{\tilde e_R}^2 - 2 m_{\tilde e_R}
\sqrt{E_+ E_-}.
\end{eqnarray}
The latter result can be used for a consistency check with the
$m_{\tilde\chi_1^0}$ measurement
from $e^+ e^-$ experiments. Assuming knowledge of
$m_{\tilde\chi_1^0}$,
we estimate the error on $m_{\tilde e_R}$ measurement as
\begin{equation}
\frac{\Delta m_{\tilde e_R}}{m_{\tilde e_R}} = \frac{1}{2}\
\frac{m_{\tilde e_R}^2 - m_{\tilde\chi_1^0}^2}
{m_{\tilde e_R}^2 + m_{\tilde\chi_1^0}^2}\
\left[ \left(\frac{\Delta E_+}{E_+}\right)^2 +
\left(\frac{\Delta E_-}{E_-}\right)^2
\right]^{\frac{1}{2}}.
\end{equation}
For a typical NLC electromagnetic calorimeter,
the single event uncertainty  on energy measurement is $\Delta E/E=
12\% /\sqrt E + 1\%$.
If there are no large systematical errors in the
measurements, the error on $\Delta m_{\tilde e_R} / m_{\tilde e_R}$
should be well below $1\%$.

\bigskip
\noindent\underline{Deducing $\tan\beta$}

The ratio of vacuum expectation values, $\tan\beta$, is of fundamental
importance in the Higgs sector. In the absence of detailed information
about Higgs boson couplings to fermions, $\tan\beta$ is difficult to
measure. Generically,
the mass splitting of the left-handed sleptons
satisfies the sum rule \cite{mr}
\begin{equation}
m_{\tilde\nu_L}^2 - m_{\tilde e_L}^2 = M_W^2 \cos2\beta \,.
\label{mtildenuL}
\end{equation}
which follows from the SU(2) structure of the left-handed
scalar partners.
Thus the $\tilde\nu_L$ and $\tilde e_L$ masses provide
an indirect measure of $\tan\beta$. From Eq.~(\ref{mtildenuL}),
the relative error in deducing $\cos2\beta$ can be determined
in terms of the measured errors
$\Delta m_{\tilde\nu_L}$ and $\Delta m_{\tilde e_L}$ as
\begin{equation}
{\left| {\Delta\cos2\beta \over \cos2\beta} \right|} =
2\ \frac{( m^2_{\tilde\nu_L} \Delta m^2_{\tilde\nu_L} +
m^2_{\tilde e_L} \Delta m^2_{\tilde e_L})^{\frac{1}{2}}}
{|m^2_{\tilde\nu_L} - m^2_{\tilde e_L}|} .
\label{Deltacos}
\end{equation}
Hence slepton mass measurements must be more accurate than
the magnitude of the sneutrino-selectron mass splitting to obtain
a significant determination of $\cos2\beta$.
The uncertainty on $\cos2\beta$ translates into an uncertainty on
$\tan\beta$ via the relation
\begin{eqnarray}
\frac{\Delta\tan\beta}{\tan\beta}
&\simeq& \frac{\tan^2\beta}{4}\ {\Delta\cos2\beta \over \cos2\beta}
\qquad {\rm for \ \tan\beta > 1},\nonumber \\
&\simeq& \frac{-1}{4\tan^2\beta}\ {\Delta\cos2\beta \over \cos2\beta}
\qquad {\rm for \ \tan\beta < 1}.
\label{tancos}
\end{eqnarray}
Thus a reasonably good sensitivity to $\tan\beta$ is obtained only when
$\tan\beta \sim 1$.
For the set of parameters under our consideration at
$\sqrt s = 500$ GeV, we find that the slepton mass
uncertainties are $\Delta m_{\tilde\nu_L}\sim 2.3$ GeV
and $\Delta m_{\tilde e_L}\sim 6$ GeV with $\pm 1\sigma$ cross section
ratio measurements.
This corresponds to an indirect
determination of $\Delta\tan\beta / \tan\beta\sim 1$ for
$\tan\beta\sim 1.8$.
%

\bigskip

\noindent\underline{Discriminating between SUSY models}

Through $e^+ e^- \to \tilde\chi^+_1\tilde\chi^-_1$
and $\tilde\chi^0_1\tilde\chi^0_2$ production
with $\tilde\chi^-_1, \tilde\chi^0_2$ subsequent decays, the NLC
will most likely
determine the MSSM parameters $M_1$, $M_2$, $\tan\beta$ and $\mu$,
although it is only possible to obtain lower bounds on
$\tan\beta$ and $\mu$ when they are large.
The information on $F_{1R}$, $V_{11}$
and $F_{2L}$, combined with the corresponding
branching fractions associated with the three $e\gamma$ processes discussed
earlier, would provide valuable additional tests of models, since the
$V_{j1}$ are functions of $\tan\beta$, $M_2$ and $\mu$,
while $F_{i(L,R)}$ depend on $\tan\beta$,  $M_1$, $M_2$
and  $\mu$.

The measurements of the slepton masses,
$m_{\tilde\nu_L}$, $m_{\tilde e_R}$ and $m_{\tilde e_L}$,
will probe soft SUSY breaking
in the scalar sector. Although SU(2) symmetry predicts
the sum rule between
the left-handed doublets in Eq.~(\ref{mtildenuL}),
there is no general relation in the MSSM
between $m_{\tilde e_R}$ and $m_{\tilde e_L}$.
In mSUGRA the left-handed and right-handed selectron masses are given
by \cite{mr,an}
\begin{eqnarray}
m^2_{\tilde e_L} & = & (m_0^{5})^2 + C_2 + \frac{1}{4} C_1 +
\left(-\frac{1}{2} + \sin^2\theta_W\right)M_Z^2\cos2\beta \nonumber\\
m^2_{\tilde e_R} & = & (m_0^{10})^2 + C_1 -
\sin^2\theta_W M_Z^2\cos2\beta
\label{selmasses}
\end{eqnarray}
where $m_0^{5}$ and $m_0^{10}$ are the scalar masses in the $\bar{5}$ and
$10$ representations of SU(5), respectively, and
\begin{eqnarray}
C_1  = \frac{2}{11}
M_1^2\left[\frac{\alpha^2(M_G)}{\alpha_1^2(M_S)}-1\right], \qquad
C_2 = \frac{3}{2}
M_2^2\left[\frac{\alpha^2(M_G)}{\alpha_2^2(M_S)}-1\right],
\label{ceq}
\end{eqnarray}
with the GUT scale $M_G \sim 10^{16}$ GeV and
the SUSY mass scale $M_S$.
In GMSB models
the selectron mass formulas are the same as above, except that
$\alpha(M_G)$ and  $m_0(M_G)$ are replaced by $\alpha(M_X)$ and
$m_0(M_X)$ where $M_X$ is the messenger mass scale.
%
%
Using the determinations of $m_{\tilde e_L}$,
$m_{\tilde e_R}$, $M_1$, $M_2$ and $\cos 2\beta$, Eqs.~(\ref{selmasses}) can
be solved for $m_0^{5}$ and $m_0^{10}$ to test
universality ($m_0^5=m_0^{10}$) of the soft supersymmetry-breaking scalar masses.

\bigskip
\noindent{\bf Summary}

We have shown that the processes
$e\gamma\to \tilde\nu\tilde\chi^-$ and $\tilde e\tilde\chi^0$
offer the opportunity to
\begin{itemize}
\item measure the selectron and sneutrino masses via the ratios
of cross sections with different initial electron and photon
polarizations,
\item estimate $\tan\beta$ through the relation in
Eqs.~(\ref{mtildenuL},\ref{tancos}),
\item test the universality of mSUGRA scalar masses at the GUT scale,
\item deduce elements of the chargino and neutralino mass
diagonalization matrices from the cross-section normalizations.
\end{itemize}
This information will be most valuable to the study of supersymmetric
unification models, especially if the thresholds for $\tilde e^+
\tilde e^-$ and $\tilde\nu \tilde\nu$ production are beyond the
kinematic reach of the Next Linear $e^+ e^-$ Collider.

\bigskip
\noindent{\bf Acknowledgments}

VB and TH would like to thank the Aspen Center for Physics for
warm hospitality during the final stages of this project.
We thank J.~Feng and X.~Tata for conversations.
This work was supported in part by the U.S.~Department of Energy
under Grants No.~DE-FG02-95ER40896 and No.~DE-FG03-91ER40674.
Further support was provided
by the University of Wisconsin Research
Committee, with funds granted by the Wisconsin Alumni Research
Foundation, and by the Davis Institute for High Energy Physics.

\newpage


\begin{references}

\bibitem{susynlc}
H. Murayama and M. E. Peskin, Ann. Rev. Nucl. Sci. {\bf 46}, 533 (1996);
Physics and Technology of the Next Linear Collider, SLAC Report 485
(submitted to 1996 Snowmass Workshop);
ECFA/DESY Linear Collider Physics Working Group, hep-ph/9705442.
%
\bibitem{fengetal}
T. Tsukamota, K. Fujii, H. Murayama, M. Yamaguchi and
Y. Okada, Phys. Rev. {\bf D51}, 3153 (1995);
J. Feng, M. E. Peskin, H. Murayama and X. Tata,
Phys. Rev. {\bf D52}, 1418 (1995);
J. L. Feng and M. J. Strassler, Phys. Rev. {\bf D51}, 4661 (1995).
%
\bibitem{msnu} A. Bartl, H. Fraas and W. Majerotto, Z. Phys.
{\bf C30}, 441 (1986).
%
\bibitem{susylhc}
H. Baer {\it et al.}, Phys. Rev. {\bf D42}, 2259 (1990);
 {\bf D50}, 4508 (1994);
I. Hinchliffe {\it et al.},
Phys. Rev. {\bf D55}, 5520 (1997).
%
\bibitem{tata}For recent comprehensive reviews and references,
see e.g. X.~Tata, Proc.\ of the {\it IX Jorge A.~Swieca Summer School},
Campos do Jord\~ao, Brazil (in press), hep-ph/9706307;
M.~Drees, KEK-TH-501, hep-ph/9611409 (1996); S. P. Martin,
hep-ph/9709356.

\bibitem{dine}M. Dine and A. E. Nelson, Phys. Rev. {\bf D48}, 1277 (1993);
M. Dine, A. E. Nelson and Y. Shirman, Phys. Rev. {\bf D51}, 1362 (1995);
M. Dine, A. E. Nelson, Y. Nir and Y. Shirman, Phys. Rev. {\bf D53}, 2658 (1996).

\bibitem{baerlhc}{ H. Baer, C.-H. Chen, F. Paige and X. Tata,
Phys. Rev. {\bf D49}, 3283 (1994).}

\bibitem{egmcollider}{H.F. Ginzburg, G.L. Kotkin, V.G. Serbo and
V.I. Telnov,
Nucl. Inst. and Meth. {\bf 205}, 47 (1983);
H.F. Ginzburg, G.L. Kotkin, S.L. Panfil, V.G. Serbo and V.I. Telnov,
Nucl. Inst. and Meth. {\bf 219}, 5 (1984).}
%
\bibitem{tanbeta}{V. Barger, M. S. Berger and P. Ohmann, Phys. Rev.
{\bf D47}, 1093 (1993); Phys. Rev. {\bf D49}, 4908 (1994);
V. Barger, M. S. Berger, P. Ohmann and R.J.N. Phillips,
Phys. Lett. {\bf B314}, 351 (1993); P. Langacker and N. Polonsky,
Phys. Rev. {\bf D50}, 2199 (1994); W. A. Bardeen, M. Carena,
S. Pokorski and C.E.M. Wagner, Phys. Lett. {\bf B320}, 110
(1994); B. Schremp, Phys. Lett. {\bf B344}, 193 (1995).}
%
\bibitem{susygut}{Analyses of supergravity mass patterns include:
G. Ross and R. G. Roberts, Nucl. Phys. {\bf B377}, 571 (1992);
R. Arnowitt and P. Nath, Phys. Rev. Lett. {\bf 69}, 725 (1992);
M. Drees and M. M. Nojiri, Nucl. Phys. {\bf B369}, 54 (1993);
S. Kelley, J. Lopez, D. Nanopoulos, H. Pois and K. Yuan,
Nucl. Phys. {\bf B398}, 3 (1993); M. Olechowski and S. Pokorski,
Nucl. Phys. {\bf B404}, 590 (1993);  V. Barger, M. Berger and
P. Ohmann, Phys. Rev. {\bf D49}, 4908, (1994); G. Kane,
C. Kolda, L. Roszkowski and J. Wells, Phys. Rev. {\bf D49}, 6173
(1994); D. J. Casta\~no, E. Piard and P. Ramond,  Phys. Rev.
{\bf D49}, 4882, (1994); W. de Boer, R. Ehret and D. Kazakov,
Z. Phys. {\bf C67}, 647 (1995); M. Noriji and X. Tata,
Phys. Rev. {\bf D50}, 2148 (1994); H. Baer, C.-H. Chen, R. Munroe,
F. Paige and X. Tata, Phys. Rev. {\bf D51}, 1046 (1995).}
%
\bibitem{dm}{See {\it e.g.} M. Drees and M. Nojiri,  Phys. Rev.
{\bf D47}, 376 (1993); R. G. Roberts and L. Roszkowski,
Phys. Lett. {\bf B309}, 329 (1993).}
%
\bibitem{robinn} {R. W. Robinett, Phys. Rev. {\bf D31}, 1657 (1985).}
%
\bibitem{selectron} F.~Cuypers, G.J.~van~Oldenberg, and R.~R\"uckl,
Nucl.\
Phys.\ {\bf B383}, 45 (1992); H.A.~K\"onig and K.A.~Peterson,
Phys.\ Lett.\
{\bf B294}, 110 (1992); D.L.~Borden, D.~Bauer and D.O.~Caldwell,
SLAC preprint SLAC-PUB-5715 (1992).
%
\bibitem{kongoto} T. Kon and A. Goto, Phys. Lett. {\bf B295}, 324 (1992).
%
\bibitem{selectron2} D.~Choudhury and F.~Cuypers, Nucl.\ Phys.
{\bf B451}, 16 (1995).
%
\bibitem{ngwu}
K.~Kiers, J.N.~Ng  and G.H.~Wu, Phys.\ Lett.\ {\bf B381}, 177 (1996).
%
\bibitem{isasugra} {H. Baer, F. Paige, S. Protpopescu, and X. Tata, in
{\it Proceedings of the Workshop On Physics at Current Accelerators and
Supercolliders}, eds. J. Hewitt, A. White and D. Zeppenfeld, Argonne
National Laboratory (1993).}
%
\bibitem{gmsusy} A. Ghosal, A. Kundu and B. Mukhopadhyaya,
Phys. Rev. {\bf D56}, 504 (1997);
S. Ambrosanio, G. D. Kribs and S. P. Martin,
Phys. Rev. {\bf D56}, 1761 (1997);
S.~Dimopoulos, S.~Thomas, and J.D.~Wells, Nucl. Phys. {\bf B488}, 39 (1997);
K.S~Babu, C.~Kolda, and F.~Wilczek, Phys. Rev. Lett. {\bf 77}, 3070 (1996).

\bibitem{baeretal}   H.~Baer, M.~Brhlik, C.~Chen, and X.~Tata, Phys. Rev.
{\bf 55}, 4463 (1997).

\bibitem{mr}{See {\it e.g.} S. Martin and P. Ramond,  Phys. Rev.
{\bf D48}, 5365 (1993).}
%
\bibitem{an}{R. Arnowitt and P. Nath, hep-ph/9708451 (1997).}

%
\end{references}
\end{document}